\documentclass[aps,prl,twocolumn,groupedaddress]{revtex4-2}

\usepackage{graphicx}
\usepackage{amsmath}
\usepackage{amssymb}
\usepackage{color}

\newcommand{\refEq}[1] {(\ref{#1})}

\newcommand{\Order}[1]{\ensuremath{O \left( #1 \right)}}
\newcommand{\Nabla}{\ensuremath{\vec{\nabla}}}
\newcommand{\romanNum}[1]{\uppercase\expandafter{\romannumeral#1}}

\begin{document}

\title{Intrinsic momentum transport driven by almost-rational surfaces in tokamak plasmas}

\author{Justin Ball}
\email[]{Justin.Ball@epfl.ch}
\author{Arnas Vol\v{c}okas}
\author{Stephan Brunner}
\affiliation{Ecole Polytechnique F\'{e}d\'{e}rale de Lausanne (EPFL), Swiss Plasma Center (SPC), CH-1015 Lausanne, Switzerland}

\date{\today}

\begin{abstract}
We demonstrate that a symmetry of the local gyrokinetic model is broken when the safety factor $q$ is almost (but not exactly) a rational number and magnetic shear is $\hat{s} \approx 0$. Tokamaks with such a $q$ profile will spontaneously rotate due to turbulent momentum transport. Nonlinear gyrokinetic simulations indicate this mechanism is significantly stronger than all other drives of intrinsic rotation. It also intrinsically transports electric current, pulling $q$ towards rational values and potentially aiding non-inductive current drive. This is likely important in the triggering of internal transport barriers.
\end{abstract}

\maketitle

Due to the toroidal symmetry of tokamak plasmas, ions and electrons are free to rotate in the toroidal direction \cite{HintonNeoclassicalRotation1985, CattoToroidalRotation1987, AbelGyrokineticsDeriv2013}. Since ions have more mass, they carry the majority of the momentum of the plasma and determine the bulk plasma rotation. Rotation is helpful as it stabilizes MHD modes, lowering the risk of disruptions \cite{BondesonToroidalStable1994, BettiRWMsoundWave1995, deVriesRotMHDStabilization1996, LiuITERrwmStabilization2004, HuRWMtrappedParticles2004, ReimerdesRWMmachineComp2006}. Additionally, strong rotation {\it gradients} can reduce turbulent transport \cite{RitzRotShearTurbSuppression1990, BurrellShearTurbStabilization1997, WaltzFlowShear1998, TerryShearedFlowSuppression2000, BarnesFlowShear2011}, a process thought to underlie many improved confinement regimes \cite{BurrellFlowShearAndHmode1994, IdaITBreview2018}.

Existing tokamaks often drive rotation externally using neutral beams, but this does not scale well to power plants \cite{LiuITERrwmStabilization2004, ParraMomentumTransitions2011}. An attractive alternative is ``intrinsic'' rotation, which is rotation that spontaneously arises from turbulent momentum transport \cite{PeetersMomTransOverview2011, ParraIntrinsicRotTheory2015, CamenenPRLExp2010, BallElongTri2016}. Turbulence enables the nested magnetic surfaces within a tokamak to push off one another and start to move. Unfortunately, intrinsic rotation is typically constrained to be weak by the symmetry properties of the tokamak. This work will demonstrate an exception.

While electrons have minimal momentum, they carry most of the toroidal plasma current. Plasma current is necessary for the tokamak equilibrium, but it is challenging to drive. Thus, the ``bootstrap'' current is helpful, which is plasma current that spontaneously arises from particle collisions \cite{BickertonBootstrapCurrent1971, GaleevBootstrapCurrent1968}. In this work, we will demonstrate intrinsic current transport by turbulence \cite{ItohTurbCurrDrive1988, ShaingCurrentDrive1988, McDevittTurbCurrentDrive2017, MaurinoThesis2025}.

To calculate momentum and current transport, we will use the local gyrokinetic (GK) model \cite{CattoLinearizedGyrokinetics1978, FriemanNonlinearGyrokinetics1982} --- a five-dimensional set of integro-differential equations thought to accurately govern turbulence in tokamaks \cite{McKeeTurbulenceScale2001}. Gyrokinetics results from a rigorous asymptotic expansion of the Fokker-Planck and Maxwell's equations in $\rho_{\ast} \equiv \rho_{i} / a \ll 1$, the ratio of the ion gyroradius $\rho_{i}$ to the tokamak minor radius $a$ \cite{AbelGyrokineticsDeriv2013}. The GK equations are usually solved in a flux tube domain \cite{BeerBallooningCoordinates1995}, which reflects the anisotropy of turbulence by being extended along magnetic field lines (i.e. $O(a)$) and narrow in the perpendicular directions (i.e. $O(\rho_{i})$). Flux tube simulations are called ``local'' as all equilibrium quantities can be Taylor expanded about $r_{0}$, the value of the minor radial coordinate $r$ at the flux tube center.

In this work, we will study momentum transport arising in certain magnetic field line topologies. ``Topology'' refers to how, if at all, the field lines close on themselves. This is determined by the safety factor $q$, the number of toroidal circuits a field line executes around the torus for each poloidal circuit. On a $q = 2$ magnetic surface, all field lines exactly close on themselves after two toroidal circuits and one poloidal circuit. On any integer $q \in \mathbb{Z}$ surface, if turbulent eddies can extend at least one poloidal turn along field lines, they will ``bite their own tails.'' This is called parallel turbulent self-interaction \cite{AjaySelfInteraction2019, BallBoundaryCond2020, VolcokasUltraLongEddies2023, VolvcokasLinearSelfInteraction2024, VolvcokasNonlinearSelfInteraction2024, StOngePhaseFactor2023}, which is necessary for the mechanism studied here. On other rational $q \in \mathbb{Q}$ surfaces (e.g. $q=3/2=1.5$), turbulent eddies can self-interact, but only if they extend multiple poloidal turns. Specifically, the denominator of the rational value corresponds to the number of poloidal turns the field line executes before closing, which is called its ``order.'' Self-interaction tends to be stronger for lower-order rational values and when magnetic shear $\hat{s} \equiv (r/q) dq/dr \ll 1$ is weak (which enables longer eddies \cite{BallBoundaryCond2020, VolcokasUltraLongEddies2023}).

An alternative topology occurs on ``almost-rational surfaces,'' where $q$ is slightly different from a low-order rational value. Even though such a {\it field line} might not bite its tail, {\it turbulent eddies} still can. This is because eddies have a finite binormal extent (i.e. the direction perpendicular to field lines but within the magnetic surface). Thus, they can encounter themselves, just somewhat misaligned. This affects the behavior of eddies and differently than when there is no binormal misalignment \cite{VolcokasUltraLongEddies2023, VolvcokasLinearSelfInteraction2024, VolvcokasNonlinearSelfInteraction2024}. Due to the anisotropy of turbulence, this requires $q$ to be within $O(\rho_{\ast})$ of a rational number (e.g. $q = 3/2 + \Order{\rho_{\ast}}$). Importantly, this does {\it not} necessarily imply the effect diminishes in larger machines because $q$ can be almost rational across the majority of the plasma \cite{JoffrinITBsplittingByRationals2002}. Lastly, a third topology occurs when $q$ is irrational (or sufficiently high-order rational) and not within $\Order{\rho_{\ast}}$ of low-order rational. In this case, eddies never encounter themselves, nor experience any self-interaction.

There is already substantial experimental evidence linking rational surfaces to momentum transport. Sawteeth \cite{HastieSawteeth1997} and other MHD instabilities \cite{MenardMagneticFluxPumping2006, PettyMagneticFluxPumping2009, ChapmanMagneticFluxPumping2010, JardinMagneticFluxPumping2015, BurckhartMagneticFluxPumping2023} redistribute current to sustain $q \gtrsim 1$, as does turbulence \cite{LiEASTturbCurrRedistribution2022, NaKSTARintrinsicCurrent2022, MaoEASTturbCurrRedistribution22023}. Moreover, many tokamaks observe rotation reversals correlated with $q \approx 1$ \cite{RiceRotFromQ2013}. Additionally, Internal Transport Barriers (ITBs) \cite{WolfITBreview2002, ConnorITBreview2004, TalaITBreview2006, IdaITBreview2018}, regions of the core exhibiting steep pressure gradients, have long been associated with turbulence stabilization by rotation shear \cite{SynakowskiRotInITB1997, ShiraiRotInITB2000, TalaRotInITB2001, HahmExBshear2002}. ITBs often appear when the safety factor profile has a region of weak magnetic shear $\hat{s} \ll 1$. Experiments on many tokamaks \cite{LopesCardozoRTPrationalITB1997, ChallisJETrationalITB2001, JoffrinITBthresholdWithRational2002, SharapovITBbeforeRational2006, JoffrinAUGrationalITB2003, AustinITBrationalDIID2006, ChungKSTARrationalITB2017, KoideJT60UrationalITB1994} and stellarators \cite{BrakelW7ASrationalITB2002, EstradaTJIIrationalITB2004, VanMilligenW7XrationalITB2018, VanMilligenTJIIrationalITB2022} found they preferentially form when low-order rational surfaces are present in the $\hat{s} \ll 1$ region \cite{IdaITBreview2018}. ITBs will follow rational surfaces radially \cite{LopesCardozoRTPrationalITB1997} and split in two if the minimum $q$ value of a reversed shear profile is lowered below a rational \cite{JoffrinITBsplittingByRationals2002}. An ITB connected to rational $q$ enabled the best-performing tokamak discharge ever \cite{FujitaRecordQshot1999}.

In this work, we will prove how the GK symmetry argument, which constrains momentum transport to be weak, is broken on almost-rational magnetic surfaces. This is a fundamental mechanism that does {\it not} require any radial inhomogeneity, can drive machine-scale net momentum, and exists in both slab and toroidal geometries. We will also present numerical simulations to indicate when this symmetry-breaking is significant.

\section{Analytic proof}

In local electrostatic gyrokinetics, one solves for the non-adiabatic portion of the turbulent distribution function $h_{s}$ and the electrostatic potential $\phi$ throughout a flux tube domain. This domain is constructed using field-aligned coordinates \cite{BeerBallooningCoordinates1995}: the magnetic surface label $x \equiv r - r_{0}$, field line label $y \equiv \left( r_{0} / q_{0} \right) \left( q (r) \chi - \zeta \right)$, position along the field line $z \equiv \chi$, parallel velocity $v_{||}$, magnetic moment $\mu$, and time $t$. The subscript $s$ indicates the species, $q_{0} = q \left( r=r_{0} \right)$, $\chi$ is a straight-field line poloidal angle, and $\zeta$ is the toroidal angle.

References \cite{PeetersMomTransSym2005, ParraUpDownSym2011, SugamaUpDownSym2011} have demonstrated that the local GK equations possess a symmetry, which is outlined in Appendix A. This symmetry implies that any solution to the GK equations, $\left( h_{s1} \left( x, y, z, v_{||}, \mu, t \right), \phi_{1} \left( x, y, z, t \right) \right)$, can be used to generate a second solution, $\left( h_{s2} \left( x, y, z, v_{||}, \mu, t \right), \phi_{2} \left( x, y, z, t \right) \right) = \left( -h_{s1} \left( -x, y, -z, -v_{||}, \mu, t \right), -\phi_{1} \left( -x, y, -z, t \right) \right)$. This second solution drives a radial flux of toroidal angular momentum $\Pi_{s2}=-\Pi_{s1}$ that cancels that of the first. This proves that the time-averaged momentum flux is $\overline{\Pi}_{s} = 0$ and, hence, so is the intrinsic rotation.

There were only four physical mechanisms known to break this symmetry \cite{PeetersMomTransOverview2011}: up-down asymmetry in the magnetic geometry, rotation, rotation shear, and higher-order effects in $\rho_{\ast} \ll 1$. Unfortunately, only one of these appears plausible in future power plants. The higher-order effects in $\rho_{\ast} \ll 1$, by definition, should weaken in larger machines. Moreover, both rotation and rotation shear will only act after something else initiates plasma rotation. Rotation driven by the remaining mechanism, up-down asymmetry, has been experimentally observed \cite{CamenenPRLExp2010} and optimized numerically \cite{BallElongTri2016, SunMomTransportLetter2025}. However, creating substantially up-down asymmetric magnetic surfaces is difficult for nearly all existing tokamaks.

Past works focused on the symmetry properties of the GK equations themselves (i.e. \eqref{eq:GKeq} and \eqref{eq:QNeq} in Appendix A). Here we will focus on the GK {\it boundary conditions}. Note that, while we will write the boundary conditions (and subsequent proof) for $\phi$, they can be formulated analogously for $h_{s}$. In the radial and binormal directions, a solution to the GK model, $\phi_{1} \left( x,y,z \right)$, must satisfy
\begin{align}
\phi_{1} \left( x + L_{x}, y, z \right) &= \phi_{1} \left( x, y, z \right) \label{eq:radialBC} \\
\phi_{1} \left( x, y + L_{y}, z \right) &= \phi_{1} \left( x, y, z \right) , \label{eq:binormalBC}
\end{align}
where $L_{x}$ and $L_{y}$ are the radial and binormal domain widths respectively. In the binormal direction, this can reflect the physical toroidal periodicity of the plasma. Indeed, the most straightforward way to faithfully model the real topology of a device is to set $L_{y}$ to the full magnetic surface width (as we will in the next section). However, radial periodicity is only justified because, in the local limit, turbulence must be {\it statistically} identical on both domain boundaries. Thus, any radial self-interaction is unphysical and must be eliminated by making the box larger, while binormal self-interaction can be physical (if $L_{y}$ is chosen appropriately \cite{VolcokasUltraLongEddies2023}).

In the parallel direction, $\phi_{1} \left( x,y,z \right)$ must satisfy the ``twist-and-shift'' boundary condition \cite{BeerBallooningCoordinates1995},
\begin{align}
\phi_{1}& \left( x, y + L_{z} \frac{r_{0}}{q_{0}} \Delta q + L_{z} \hat{s} x, z + L_{z} \right) = \phi_{1} \left( x, y, z \right) , \label{eq:parallelBC}
\end{align}
where $L_{z} = 2 \pi N_{pol}$ and $N_{pol} \in \mathbb{Z}$ quantifies the length of the domain in number of poloidal turns. It differs from simple periodicity due to two terms. The first represents the binormal {\it shift} of the center of the flux tube after $N_{pol}$ poloidal turns due to the value of $q_{0}$. The second expresses the {\it twist} created by $\hat{s}$ (i.e. different radial locations will have different shifts due to the varying $q$). In the first term, we have already repeatedly applied \eqref{eq:binormalBC} to shift field lines by a discrete binormal distance $\pm L_{y}$, so only the modulus $\Delta q \equiv q_{0} - \delta q ~ \lfloor q_{0} / \delta q \rfloor$ remains. Here $\lfloor \ldots \rfloor$ is the floor function and $\delta q \equiv q_{0} L_{y} / ( r_{0} L_{z} )$ is the effective shift in $q$ accomplished by each application of binormal periodicity. {\it We claim the term containing $\Delta q$ breaks the symmetry of the local GK model when $\hat{s} = 0$}.

The proof is composed of two parts. The {\it first part} starts by assuming we have a solution to the GK model, $\phi_{1} \left( x, y, z \right)$, which we know satisfies the boundary conditions of \refEq{eq:radialBC}, \refEq{eq:binormalBC}, and \refEq{eq:parallelBC} as well as the gyrokinetic and quasineutrality equations (i.e. \refEq{eq:GKeq} and \refEq{eq:QNeq} in Appendix A). From this, we postulate a second solution $\phi_{2} \left( x, y, z \right) \equiv - \phi_{1} \left( -x, y, -z \right)$ to the same GK model. We know from references \cite{PeetersMomTransSym2005, ParraUpDownSym2011, SugamaUpDownSym2011} that $\phi_{2}$ solves the gyrokinetic and quasineutrality equations (assuming the above symmetry-breaking mechanisms are absent). However, we must also prove $\phi_{2}$ satisfies the boundary conditions --- in particular, its parallel boundary condition
\begin{align}
	\phi_{2}& \left( x, y + L_{z} \frac{r_{0}}{q_{0}} \Delta q + L_{z} \hat{s} x, z + L_{z} \right) = \phi_{2} \left( x, y, z \right) . \label{eq:parallelBC2}
\end{align}
For the {\it radial} boundary condition, this is done by substituting the definition of $\phi_{2}$ and making the coordinate transformation $\left( X,Y,Z \right) = \left( -x-L_{x}, y, -z \right)$, yielding $\phi_{1} \left( X,Y,Z \right) = \phi_{1} \left( X + L_{x}, Y, Z \right)$. Since \refEq{eq:radialBC} holds for any value of $\left( x, y, z \right)$, we know that this equation will be true for all values $\left( X,Y,Z \right)$, which implies $\phi_{2}$ satisfies its radial boundary condition. For the binormal boundary condition, we repeat the same process using $\left( X,Y,Z \right) = \left( -x, y, -z \right)$. For the parallel boundary condition, we start by substituting the definition of $\phi_{2}$ into \refEq{eq:parallelBC2} to get 
\begin{align}
	\phi_{1}& \left( -x, y + L_{z} \frac{r_{0}}{q_{0}} \Delta q + L_{z} \hat{s} x, -z - L_{z} \right) = \phi_{1} \left( -x, y, -z \right) , \label{eq:parallelBC2mod}
\end{align}
which we must transform into a form equivalent to \refEq{eq:parallelBC}. For the radial coordinate, we must substitute $X = - x$. For the parallel coordinate, the only choice is $Z = - z - L_{z}$, which also implies that we must match the left side of \refEq{eq:parallelBC2mod} with the right side of \refEq{eq:parallelBC} and the right side of \refEq{eq:parallelBC2mod} with the left side of \refEq{eq:parallelBC}. For the binormal coordinate, doing the first task requires $Y = y + L_{z} r_{0} \Delta q / q_{0} + L_{z} \hat{s} x$, while the second requires $Y = y - L_{z} r_{0} \Delta q / q_{0} + L_{z} \hat{s} x$. These two requirements are incompatible, so the task is impossible. However, if $\Delta q = 0$, we can substitute $Y = y + L_{z} \hat{s} x$ and show $\phi_{2}$ satisfies \refEq{eq:parallelBC2}. Thus, if $\Delta q = 0$, the standard symmetry argument works, proving that $\Pi_{s2}=-\Pi_{s1}$ and the time-averaged momentum flux is $\overline{\Pi}_{s} = 0$. If $\Delta q \neq 0$, the standard symmetry argument is broken.

In the {\it second part} of the proof, we consider a GK model that is identical to that of $\phi_{1}$, but without the $\Delta q$ term in \refEq{eq:parallelBC}. We postulate this new system is solved by $\phi_{\mathbb{Q}} \left( x, y, z \right) \equiv \phi_{1} \left( x - x_{\mathbb{Q}}, y, z \right)$, where $x_{\mathbb{Q}} \equiv ( r_{0} / \hat{s} ) \Delta q / q_{0}$. Importantly, this assumes $\hat{s} \neq 0$. Since we already know $\phi_{1}$ is a solution and $x$ does not appear explicitly in the gyrokinetic nor quasineutrality equations, $\phi_{\mathbb{Q}}$ must solve them as well. Next, we take all three boundary conditions for $\phi_{\mathbb{Q}}$, substitute the definition of $\phi_{\mathbb{Q}}$, and then $\left( X,Y,Z \right) = \left( x - x_{\mathbb{Q}}, y, z \right)$. This yields equations identical in form to \refEq{eq:radialBC}-\refEq{eq:parallelBC}. Thus, $\phi_{\mathbb{Q}}$ is a valid solution to its GK model. Since its model has no $\Delta q$ term, the standard symmetry argument proves $\overline{\Pi}_{s\mathbb{Q}} = 0$. Moreover, since $\phi_{1}$ and $\phi_{\mathbb{Q}}$ only differ by a translation in $x$ and the momentum flux (given by \eqref{eq:momFlux} in Appendix A) is averaged over $x$, we know that $\Pi_{s1} = \Pi_{s\mathbb{Q}}$. Thus, the time-averaged momentum flux is still $\overline{\Pi}_{s} = 0$ when $\Delta q \neq 0$, as long as $\hat{s} \neq 0$.

In summary, if $\Delta q = 0$, we can apply the standard symmetry argument to show that $\overline{\Pi}_{s} = 0$, regardless of $\hat{s}$. If $\Delta q \neq 0$ but $\hat{s} \neq 0$, we can radially translate the  solution to eliminate $\Delta q$ (without affecting $\Pi_{s}$) and then apply the standard symmetry argument to show that $\overline{\Pi}_{s} = 0$. However, if $\Delta q \neq 0$ and $\hat{s} = 0$, translating the solution does not remove $\Delta q$, so we cannot apply the symmetry argument and $\overline{\Pi}_{s}$ is permitted to be non-zero. Note that this proof shows which combinations of $q_{0}$, $L_{y}$, and $L_{z}$ break the symmetry, but not which are physically possible. However, a straightforward way to ensure this is to set $L_{z} = 2 \pi$ and $L_{y}$ to the full binormal width of the magnetic surface.

To intuitively understand the symmetry-breaking at $\hat{s}=0$, consider e.g. ion drift waves, which possess a velocity $\vec{v}_{\ast i} \propto \vec{B} \times \Nabla p_{i}$ in the $-y$ direction. If $\Delta q > 0$, particles with $v_{||} > 0$ will also move the $-y$ direction whenever they pass through the parallel boundary, allowing them to resonate with the drift waves through their parallel motion. However, particles with $v_{||} < 0$ will travel through the parallel boundary in the opposite direction, causing \refEq{eq:parallelBC} to shift them in the $+y$ direction and preventing them from resonating with the drift wave. Thus, particles traveling in opposite directions will be transported differently thereby breaking the symmetry, but only when $\Delta q$ is small enough that thermal particles resonate with the instability.

In a flux tube, this symmetry-breaking formally exists only for exactly $\hat{s} = 0$, but in real machines it will persist for small but finite $\hat{s} \approx 0$. This is because, when $\hat{s}$ is finite, the boundary conditions force the flux tube to span the radial distance between lowest-order rational surfaces, assuming a Taylor-expanded linear $q$ profile \cite{BeerBallooningCoordinates1995, BallBoundaryCond2020}. This means the domain will symmetrically include both positive and negative $\Delta q$, which locally drive values of $\overline{\Pi}_{s}$ that cancel (as our proof requires). However, at low $\hat{s}$, the radial domain can be larger than the physical machine and often poorly approximates the actual $q$ profile. In reality, machines are not constrained to symmetrically include the two canceling regions of positive and negative $\Delta q$.

\section{Numerical simulations}

We will now use the GENE code \cite{JenkoGENE2000, GermaschewskiGPUgene2021} to numerically solve the local GK model and calculate the momentum transport driven by almost-rational surfaces. While the analytic proof of symmetry-breaking is an abstract mathematical consequence of the boundary conditions, here we seek to accurately model its impact in a hypothetical $q_{0}$ scan with $\hat{s} = 0$ in a TCV-sized tokamak \cite{HofmannTCVOverview1994}. This could be accomplished experimentally by creating a reversed shear safety factor profile and then performing a current ramp to change $q_{0}$ at the location where $\hat{s} = 0$. To ensure the physical parallel and binormal self-interaction, we set the domain to the full magnetic surface \cite{BallBoundaryCond2020} and varied $\Delta q$ consistently with $q_{0}$. Appendix B explains the details of how this is done and the simulation parameters/normalizations.

Figure \ref{fig:numResults}(a) shows the total heat flux from ions and electrons. It varies dramatically and exhibits fine structure around rational surfaces up to 10\textsuperscript{th} order \cite{VolcokasUltraLongEddies2023, StOngePhaseFactor2023}.

Figure \ref{fig:numResults}(b) shows the ion momentum flux. We see that it is zero at $q_{0} = 1$ and $q_{0} = 2$ (where $\Delta q = 0$) as well as at other low-order rational values like $q_{0} = 1.5$ (as eddies still exactly bite their own tails after few poloidal turns). Nearby $q_{0} = 1$, the momentum flux driven by almost-rational surfaces is strong, almost triple the maximum ever obtained from optimized up-down asymmetry \cite{BallElongTri2016}. In comparison, the peaks around $q_{0} = 2$ are lower, broader, and modulated by neighboring higher-order rational values. We also ran a few simulations with finite rotation shear $\omega_{ExB} \equiv \left( r_{0}/q_{0} \right) d \Omega_{\zeta}/dr$ to achieve $\overline{\Pi}_{s} = 0$  (red), which predicts the $\omega_{ExB}$ that would arise in a steady-state experiment. This yielded substantial values of intrinsic rotation shear that were sufficient to affect the heat flux.

Figure \ref{fig:numResults}(c) shows the electron momentum flux, which represents the transport of electric current. We see peaks at rational surfaces (up to 7\textsuperscript{th} order), which are narrower than those in $\overline{\Pi}_{i}$. This is a result of the faster electron parallel streaming, which leads them to resonate with the drift waves for a smaller $\Delta q$. Moreover, all peaks have identical parity around the rational value, which ensures that the current transport always modifies the $q$ profile to pull it closer to the rational value (as observed in \cite{VolcokasSafetyFactorFlattening2024, DiGiannataleSafetyFactorFlattening2025}). The height of these peaks is more than $10$ times typical values from up-down asymmetry and can compete with resistive dissipation in the core (see Appendix B).

We find that $\overline{\Pi}_{i}$ and $\overline{\Pi}_{e}$ change sign if $\Delta q$ changes sign or the direction of the plasma current is flipped, but {\it not} if the toroidal magnetic field direction is flipped. This ensures turbulence always pushes the $q$ profile towards rational values. Given that $q$ must be within $\Order{\rho_{\ast}}$ of the rational value for turbulent self-interaction to occur, the width of the structures in figure \ref{fig:numResults} will narrow in larger devices, while their height remains similar \cite{VolvcokasNonlinearSelfInteraction2024}.

\section{Conclusions}

We have demonstrated that intrinsic momentum transport is driven when $\hat{s} \approx 0$ and the safety factor has a value almost (but not exactly) low-order rational. This means that a tokamak with, e.g., $q \approx 1.03$ over a substantial fraction of its minor radius will rotate without external momentum injection. Additionally, almost-rational $q$ values intrinsically transport electric current that pulls the profile towards rational and makes it ``sticky,'' i.e. a safety factor profile that is flat and rational will tend to redistribute current to keep it that way. This could aid in non-inductive current drive. We expect these effects to become more important in a reactor, as strong core alpha heating will strengthen turbulence relative to resistivity. Lastly, irrespective of momentum transport, almost-rational $q$ values dramatically reduce turbulence.

We believe these sharp changes in turbulent heat, momentum, and current transport are key to the formation of ITBs in tokamaks and stellarators \cite{FujitaRecordQshot1999, ChaudharyW7Xitb2024}, but they also represent a challenge for plasma control systems. While disruptions have long been associated with rational $q$ \cite{FujitaRecordQshot1999}, this is typically attributed to enhanced MHD instability. Our work suggests that turbulence may also be to blame. If the sharp changes in transport can be incorporated into control systems, it may reveal flat rational $q$ profiles are more stable than expected.

\begin{acknowledgments}
{\it Acknowledgments} --- The authors would like to thank F. Parra, P. Ivanov, H. Sun, and G. Di Giannatale for useful discussions pertaining to this work, as well as T. G\"{o}rler and G. Merlo for their assistance with the GPU-enabled version of GENE. 
This work has been carried out within the framework of the EUROfusion Consortium, partially funded by the European Union via the Euratom Research and Training Programme (Grant Agreement No 101052200 — EUROfusion). The Swiss contribution to this work has been funded in part by the Swiss State Secretariat for Education, Research and Innovation (SERI). Views and opinions expressed are however those of the author(s) only and do not necessarily reflect those of the European Union, the European Commission or SERI. Neither the European Union nor the European Commission nor SERI can be held responsible for them.
This work was supported by a grant from the Swiss National Supercomputing Centre (CSCS) under project ID s1249, lp34, and lp06.
\end{acknowledgments}

\appendix
\section{Appendix A: Symmetry of GK equations}
\label{app:symBreaking}

The local GK model (in the absence of electromagnetic effects, collisions, and equilibrium rotation) is composed of the GK equation \cite{ParraUpDownSym2011}
\begin{align}
\frac{\partial \textcolor{red}{h_{s}}}{\partial t} & + \textcolor{red}{v_{||}} \hat{b} \cdot \Nabla z \frac{\partial \textcolor{red}{h_{s}}}{\partial \textcolor{red}{z}} + \textcolor{red}{\vec{v}_{Ds} \cdot \Nabla x} \frac{\partial \textcolor{red}{h_{s}}}{\partial \textcolor{red}{x}} + \vec{v}_{Ds} \cdot \Nabla y \frac{\partial \textcolor{red}{h_{s}}}{\partial y} \nonumber \\
&+ \textcolor{red}{a_{s||}} \frac{\partial \textcolor{red}{h_{s}}}{\partial \textcolor{red}{v_{||}}} + \mathcal{C} \left( \frac{\partial \textcolor{red}{\left\langle \phi \right\rangle_{\varphi}}}{\partial \textcolor{red}{x}} \frac{\partial \textcolor{red}{h_{s}}}{\partial y} - \frac{\partial \textcolor{red}{\left\langle \phi \right\rangle_{\varphi}}}{\partial y} \frac{\partial \textcolor{red}{h_{s}}}{\partial \textcolor{red}{x}} \right) \label{eq:GKeq} \\
&= \frac{Z_{s} e F_{Ms}}{T_{s}}\frac{\partial \textcolor{red}{\left\langle \phi \right\rangle_{\varphi}}}{\partial t} + \mathcal{C} \frac{\partial \textcolor{red}{\left\langle \phi \right\rangle_{\varphi}}}{\partial y} \frac{d F_{M s}}{d x} , \nonumber
\end{align}
the quasineutrality equation (i.e. Gauss's law)
\begin{align}
\sum_{s} Z_{s} e B \int d v_{||} \int d \mu \oint d \varphi &~ \textcolor{red}{h_{s}} = \sum_{s} \frac{Z_{s}^2 e^2 n_{s}}{T_{s}} \textcolor{red}{\phi} , \label{eq:QNeq}
\end{align}
and the boundary conditions of \eqref{eq:radialBC}-\eqref{eq:parallelBC}. Here $\hat{b} \equiv \vec{B}/B$, $\vec{v}_{Ds}$ is the magnetic drift velocity, $a_{s||}$ is the parallel acceleration, $\mathcal{C} \equiv (r_{0} / q_{0})  dx/d \psi$ is a geometrical constant (which equals $\mathcal{C} = 1$ for circular, large aspect ratio magnetic surfaces), $\left\langle \ldots \right\rangle_{\varphi}$ is the gyroaverage (at fixed guiding center position), $Z_{s}$ is the particle charge number, $e$ is the elementary charge, $F_{Ms}$ is the background Maxwellian distribution function, $T_{s}$ is the background temperature, $B$ is the magnetic field strength, $\mu \equiv v_{\perp}^2 / (2 B)$, $\oint d \varphi \left( \ldots \right)$ signifies an integral over gyroangle $\varphi$ (at fixed particle position), $n_{s}$ is the density, $\psi$ is the poloidal magnetic flux, and $v_{\perp}$ is the perpendicular speed.

Equations \refEq{eq:GKeq} and \refEq{eq:QNeq} possesses a symmetry \cite{PeetersMomTransSym2005, ParraUpDownSym2011, SugamaUpDownSym2011}. If one simultaneously changes the sign of the radial coordinate $x \rightarrow -x$, the parallel coordinate $z \rightarrow -z$, and the parallel velocity coordinate $v_{||} \rightarrow -v_{||}$ as well as $h_{s} \rightarrow -h_{s}$ and $\phi \rightarrow -\phi$, the equations remain unchanged. To elucidate this, in \refEq{eq:GKeq} and \refEq{eq:QNeq}, we mark in red the quantities that change sign under this transformation. Importantly, every term has an odd number of red quantities, meaning that all of the negative signs cancel out. Though not shown here, this symmetry holds even when including electromagnetic fields and collisions \cite{ParraUpDownSym2011, SugamaUpDownSym2011}.

This symmetry means we can take any solution to \refEq{eq:GKeq} and \refEq{eq:QNeq}, $\left( h_{s1} \left( x, y, z, v_{||}, \mu, t \right), \phi_{1} \left( x, y, z, t \right) \right)$, and generate a second solution, $\left( h_{s2} \left( x, y, z, v_{||}, \mu, t \right), \phi_{2} \left( x, y, z, t \right) \right) = \left( -h_{s1} \left( -x, y, -z, -v_{||}, \mu, t \right), -\phi_{1} \left( -x, y, -z, t \right) \right)$. This is relevant to momentum transport because the radial flux of toroidal angular momentum due to turbulence can be written as \cite{ParraUpDownSym2011}
\begin{align}
\Pi_{s} &= - \left\langle \mathcal{C} B \int d v_{||} \int d \mu \oint d \varphi ~ m_{s} R^2 \textcolor{red}{\vec{v}} \cdot \Nabla \zeta \textcolor{red}{h_{s}} \frac{\partial \textcolor{red}{\phi}}{\partial y} \right\rangle_{x,y,z} , \label{eq:momFlux}
\end{align}
where $m_{s}$ is the particle mass, $R$ is the major radial coordinate, $\vec{v}$ is the velocity, and $\left\langle \ldots \right\rangle_{x,y,z}$ is a volume average over the whole flux tube domain. Thus, we see the second solution has a momentum flux $\Pi_{s2}=-\Pi_{s1}$, which exactly cancels that of the first solution. Given that turbulence is chaotic, if we wait longer than the turbulent correlation time, both solutions are statistically likely to appear and cancel. This proves that $\overline{\Pi}_{s} = 0$, where $\overline{\left( \ldots \right)}$ is a sufficiently long time average. Since turbulence cannot move momentum around, this argument also implies that the intrinsic rotation will be zero. This also prevents turbulence from redistributing electric current as the radial flux of current is $(Z_{s} e / m_{s} ) \overline{\Pi}_{s}$.

There are, however, four known mechanisms that break the symmetry of \refEq{eq:GKeq} and \refEq{eq:QNeq} \cite{PeetersMomTransOverview2011}. First, if the equilibrium is up-down asymmetric (i.e. the magnetic surfaces do {\it not} have reflectional symmetry about the midplane), than the various geometrical coefficients no longer have a defined parity in $z$. For the symmetry to work, we need, for example, $\hat{b} \cdot \Nabla z$ to be even in $z \rightarrow -z$ and $\vec{v}_{Ds} \cdot \Nabla x$ to be odd. Up-down asymmetric magnetic surfaces result in geometric coefficients that are neither even nor odd, breaking the overall symmetry. Second and third, both equilibrium toroidal rotation and equilibrium toroidal rotation shear add new terms to \refEq{eq:GKeq} that do not respect the symmetry. Lastly, there are many effects that are higher order in the $\rho_{\ast} \ll 1$ expansion of gyrokinetics that break the symmetry \cite{ParraIntrinsicRotTheory2015} (e.g. turbulent intensity gradient \cite{ItohTurbCurrDrive1988, GurcanTurbIntensityGradCurrDrive2010, LuRotTurbIntensityGrad2015}, profile shearing \cite{WaltzMomRadialProfileVar2011, SinghMomRadialProfileVar2014}, the parallel nonlinearity \cite{McDevittRotParallelNonlinearity2009}, non-Maxwellian distribution functions \cite{BarnesRotNonMaxwellDist2013}). As discussed in the main text, all these four options have significant limitations that inhibit driving fast rotation.

\section{Appendix B: Simulation parameters}
\label{app:simParams}

\begin{table*}
	\caption{\label{tab:simParams}
		The simulation parameters used for figure \ref{fig:numResults}, which employed kinetic electrons and circular magnetic surfaces (specified using the Miller model \cite{MillerGeometry1998}). Note that all grids are equally spaced, $R_{0}$ is the tokamak major radius, $\rho_{thi} \equiv \sqrt{m_{i} T_{i}}/(e B_{0})$, $v_{ths} \equiv \sqrt{2T_{s}/m_{s}}$ is the thermal velocity, and $B_{0}$ is the toroidal magnetic field strength at $R=R_{0}$.
	}
	\begin{ruledtabular}
		\begin{tabular}{cc|cc}
			Parameter & Value & Parameter & Value \\
			\colrule
			\hline
			Minor radius of flux tube, $r_{0}/R_{0}$ & $0.18$ & Magnetic shear, $\hat{s}$ & $0.0$ \\
			Ion temperature gradient, $R_{0}/L_{Ti}$ & $6.96$ & Electron temperature gradient, $R_{0}/L_{Te}$ & $3.48$ \\
			Density gradient, $R_{0}/L_{n}$ & $2.22$ & Plasma $\beta$ & $10^{-5}$ \\
			Ion-electron mass ratio, $m_{i}/m_{e}$ & $3671$ & Ion-electron temperature ratio, $T_{i}/T_{e}$ & $1.0$ \\
			4\textsuperscript{th} order $\chi$ hyperdiffusion \cite{PueschelHyperDiff2010}, $\epsilon_{\chi}$ & $0.2$ & 4\textsuperscript{th} order $v_{||}$ hyperdiffusion \cite{PueschelHyperDiff2010}, $\epsilon_{v||}$ & $0.2$ \\
			\hline
			$x / \rho_{thi}$ range, $[ 0, L_{x} / \rho_{thi} )$ & $[ 0, 200 )$ & Number of $x$ grid points, $N_{x}$ & $256$ \\
			$y / \rho_{thi}$ range, $[ 0, L_{y} / \rho_{thi} )$ & $[ 0, 250/q_{0})$ & Number of $y$ grid points, $N_{y}$ & $\lfloor 256/q_{0} \rfloor$ \\
			$z$ range & $[ -\pi, \pi )$ & Number of $z$ points, $N_{z}$ & $16$ \\
			$v_{||} / v_{ths}$ range & $[ - 3, 3 ]$ & Number of $v_{||}$ grid points, $N_{v||}$ & $48$ \\
			$\sqrt{\mu / (T_{s} / (m_{s} B_{0} ))}$ range & $( 0, 2.72 ]$ & Number of $\sqrt{\mu}$ grid points, $N_{\mu}$ & $8$ \\
		\end{tabular}
	\end{ruledtabular}
\end{table*}

\begin{turnpage}
	\begin{figure*}
		\centering
		\includegraphics[width=1.0\textheight]{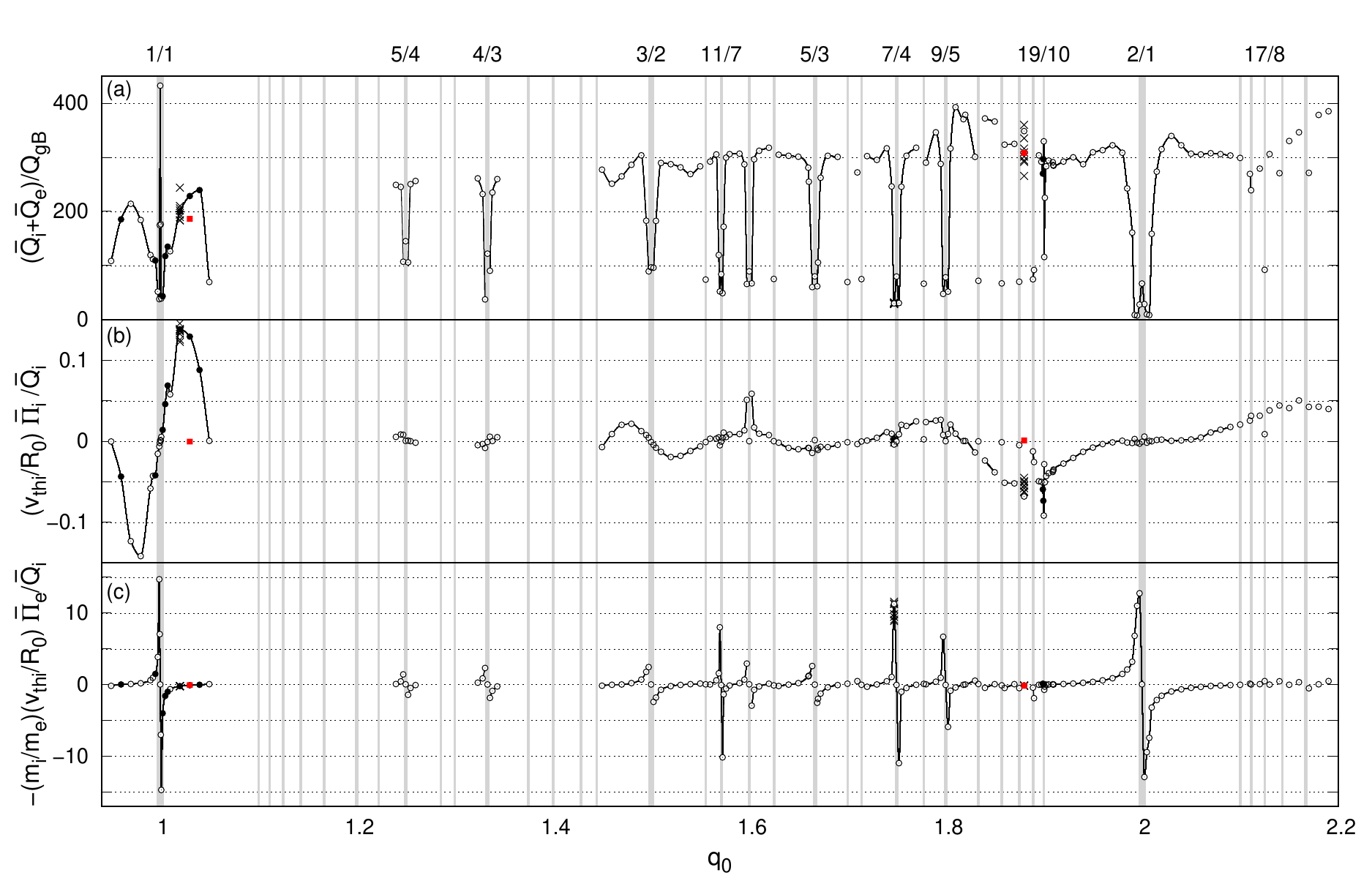}
		\caption{The (a) total heat flux, (b) intrinsic ion momentum flux, and (c) intrinsic electric current flux as a function of safety factor from $\sim 200$ nonlinear GK simulations with $\hat{s}=0$. Background toroidal flow shear is not included (black circles) except at $q=1.03$ and $q=1.88$ (red squares), where it is $\omega_{ExB}=-0.11 v_{thi}/R_{0}$ and $\omega_{ExB}=0.03 v_{thi}/R_{0}$ respectively. Filled circles indicate simulations exhibiting ``bursty'' heat flux time traces \cite{VolvcokasNonlinearSelfInteraction2024}. A line connects neighboring points where we believe the resolution is sufficient to capture the variation in all three plots.}
		\label{fig:numResults}
	\end{figure*}
\end{turnpage}

Table \ref{tab:simParams} shows the parameters used for figure \ref{fig:numResults}. The gradients were inspired by the Cyclone Base Case \cite{DimitsCBC2000}, with a reduced $R_{0} / L_{Te}$ to stabilize the electron temperature gradient (ETG) modes. The safety factor $q_{0}$ was scanned while keeping $\hat{s} = 0$. It was necessary to treat electrons kinetically to accurately capture parallel self-interaction \cite{DominskiCorrugations2015}. Extensive resolution studies were performed, in which the number of grid points and the domain width (at constant grid spacing) in each dimension were individually doubled. Results are shown for $q_{0} = \left\{ 1.02, 1.7475, 1.88 \right\}$ in figure \ref{fig:numResults}, indicated by the eight black crosses for each $q_{0}$. Additionally, the parallel boundary condition has been extensively tested \cite{VolvcokasLinearSelfInteraction2024} and the momentum transport from almost-rational surfaces was successfully benchmarked against an analytic solution for linear modes in a slab geometry. Note that, since $\hat{s} = 0$, the radial box width $L_{x}$ could be freely chosen and was not constrained by the typical quantization condition $L_{x} = N L_{y} / (2 \pi N_{pol} \left| \hat{s} \right| )$, where $N \in \mathbb{Z}$. When $q_{0}$ is extremely close to (but not at) rational values, we find that large values of $L_{x}$ are sometimes necessary to resolve $\overline{\Pi}_{e}$, though $y$ grid spacings larger than the parallel boundary condition binormal offset appear acceptable. In the resolution study, $L_{y}$ and $L_{z}$ were not doubled as they are set to the physical values needed to model the full magnetic surface. This is essential to ensure that the physical field line connection properties are faithfully replicated by the simulation domain \cite{BallBoundaryCond2020}. To do so, we set the length of the domain to $N_{pol} = 1$ poloidal turn and the toroidal extent $L_{\zeta}$ to the full physical value $L_{\zeta} = 2 \pi$. From the definition of the binormal coordinate $y$, we see that (when holding $x$ and $z$ constant) this corresponds to $L_{y} = 2 \pi r_{0}/q_{0}$. Thus, in addition to the standard variation of the geometric coefficients in \refEq{eq:GKeq} with $q_{0}$, we also changed the binormal box size (keeping the grid spacing constant) as we scanned $q_{0}$. We note that, when normalizing $L_{y} = 2 \pi r_{0}/q_{0}$, the left side is expressed in terms of the ion thermal gyroradius $\rho_{thi}$, while the right side is in terms of the minor radius $a$. This results in a factor of $\rho_{\ast} \equiv \rho_{thi} / a$, which was chosen to be $\rho_{\ast} \approx 1 / 75$ to be similar to TCV \cite{CamenenPRLExp2010} for its aspect ratio of $R_{0}/a \approx 3$. The value of $\rho_{\ast}$ determines the toroidal circumference of the magnetic surface, expressed in number of ion gyroradii. We stress the role of $\rho_{\ast}$ and its value in the simulations as we expect it to significantly affect the width (but not the height \cite{VolvcokasNonlinearSelfInteraction2024}) of the structures around rational values in figure \ref{fig:numResults}. They should narrow for smaller $\rho_{\ast}$ (i.e. larger machines) because a similar-sized turbulent eddy biting its own tail with the same binormal offset (in number of $\rho_{thi}$) will correspond to a smaller offset in the safety factor value. The formula $\Delta y / \rho_{thi} = 2 \pi n \left( r_{0} / a \right) \left( 1 / \rho_{\ast} \right) \left( \Delta q / q_{0} \right)$ relates the offset from a rational $q$ value to the binormal offset in number of ion gyroradii, where $n \in \mathbb{Z}$ is the order of the nearby rational.

To quantify the heat flux in our simulations, we used a standard GENE gyroBohm normalization $Q_{gB} \equiv n_{e} T_{i} \sqrt{T_{i}/m_{i}} \left( \rho_{thi} / R_{0} \right)^2$. In our simulations, the heat flux was dominantly carried by ions, indicating Ion Temperature Gradient (ITG)-driven turbulence. To quantify momentum transport, we normalized the ion toroidal angular momentum flux to the ion heat flux according to $\left( v_{thi} / R_{0} \right) \overline{\Pi}_{i} / \overline{Q}_{i}$. This metric allows one to estimate the rotation gradient that will arise from a given temperature gradient \cite{BallElongTri2016}, which is then easy to compare with experiment \cite{BallMomUpDownAsym2014}. The electron momentum flux, $- \left( m_{i} / m_{e} \right) \left( v_{thi} / R_{0} \right) \overline{\Pi}_{e} / \overline{Q}_{i}$, was normalized with a negative sign and the mass ratio in order to convert it into a flux of electric current. This quantity can be compared with $\left( v_{thi} / R_{0} \right) \overline{\Pi}_{i} / \overline{Q}_{i}$ to verify that electrons carry the vast majority of the current, but a negligible amount of momentum.

To get a sense for the experimental significance of the turbulent flux of current, we can compare it to the rate at which resistivity dissipates the plasma current \cite{McDevittTurbCurrentDrive2017}. A radial flux of current is certainly important if it can bring enough current through a magnetic surface to replenish the current that is lost within the surface due to resistivity. Thus, guided by the Braginskii momentum balance equation \cite{WessonTokamaks2004}, we will compare the toroidal angular momentum flux to the integral of the friction force $R_{e}$ according to $\Pi_{e} \sim \int_{0}^{r_{0}} dr R_{0} R_{e}$, where the factor of $R_{0}$ is needed to convert force into torque. Approximating the integral and normalizing in the same way as $\Pi_{e}$, we find the metric $\left( m_{i} / m_{e} \right) \left( v_{thi} / R_{0} \right) \left( r_{0} R_{0} R_{e} \right) / Q_{heat}$. This can be calculated for various experiments and compared with figure \ref{fig:numResults}(c). Here $Q_{heat}$ is the total heating power divided by the area of the $r=r_{0}$ magnetic surface, the friction force experienced by electrons is $R_{e} = n_{e} e \eta j$, $\eta$ is the Spitzer resistivity, and $j$ is the plasma current density. This metric indicates that turbulent current transport will be most important close to the magnetic axis in hot, low density plasmas with strong core heating.

For a KSTAR discharge exhibiting substantial anomalous current around $r_{0}/a = 0.3$ \cite{NaKSTARintrinsicCurrent2022}, we estimate $\left( m_{i} / m_{e} \right) \left( v_{thi} / R_{0} \right) \left( r_{0} R_{0} R_{e} \right) / Q_{heat} \approx 10$. Similarly, we find the metric is $\approx 3$ and $\approx 10$  at $r_{0}/a = 0.3$ in anomalous current-drive experiments on EAST \cite{LiEASTturbCurrRedistribution2022} and ASDEX-U \cite{BurckhartMagneticFluxPumping2023}, respectively. Standard scenarios in ITER \cite{AymarITERSummary2001} and EU-DEMO \cite{KembletonEUDEMO2022} both give an expected metric of $\approx 14$ at $r_{0}/a = 0.3$. As figure \ref{fig:numResults}(c) displays comparable values, this suggests that turbulent current transport from almost-rational surfaces could entirely sustain substantial plasma current in the core of many tokamaks.

Lastly, the metric goes to zero in ``current hole'' experiments \cite{FujitaJT60UcurrentHole2001, HawkesJETcurrentHole2001}, in which the on-axis current density is zero, clearly indicating that turbulent current drive could be important. Moreover, current holes (along with X-points) maximize the effect of self-interaction as the field lines close on themselves after one toroidal turn and zero poloidal turns.

\bibliography{references}

\clearpage

\global\pdfpageattr\expandafter{\the\pdfpageattr/Rotate 90}

\end{document}